\documentclass[reprint,amsmath,amssymb,aps,prl,showpacs,floatfix]{revtex4-1}

\usepackage{graphicx}
\usepackage{dcolumn}
\usepackage{bm}
\usepackage{hyperref}
\usepackage{subfigure}

\begin{document}

%\preprint{APS/123-QED}

\title{Maximal tripartite entanglement between singlet-triplet qubits in quantum dots}

\author{Tuukka Hiltunen}
\author{Ari Harju}
%\author{Anybody Else?$^1$}

\affiliation{%
COMP Centre of Excellence, Department of Applied Physics, Aalto University, Helsinki, Finland
}%

\date{\today}

\begin{abstract}
We propose an efficient three-qubit gate of singlet-triplet states in quantum dots based on the capacitative coupling.
This scheme can be used to generate the maximally entangled Greenberger-Horne-Zeilinger state with one simple and fast gate operation.
Our simulations using a realistic microscopic model combined with our detailed
analysis for the gate operation can be used to extract the actual experimental pulse sequence needed to realize this.
%\begin{description} %\item[Usage] %Secondary publications and information retrieval purposes.
%\item[PACS numbers] %May be entered using the \verb+\pacs{#1}+
%\end{description}
\end{abstract}

 %CHECK!
%\pacs{Valid PACS appear here}
\maketitle

\noindent 
Entanglement is an essential resource in quantum information technology. It is exploited in for example
quantum teleportation, and entangled states are at the heart of all quantum computation \cite{ent}. 
Entanglement is, however, fragile to the environment, and it cannot generally be increased with local operations if the
involved parties are not in direct contact \cite{ent,dur}. The development of efficient methods for generating highly entangled states is thus an important
task.

In contrast to the bipartite case, where all maximally entangled states are equivalent up to local operations \cite{dur},
genuine tripartite entanglement exhibits two different classes,
the W-states \cite{ent,dur} and 
the Greenberger-Horne-Zeilinger (GHZ) states \cite{greenberger}.
The GHZ-states, represented by $|GHZ^{\pm}\rangle=(|000\rangle\pm|111\rangle)/\sqrt{2}$,
are especially interesting, as they
exhibit the strongest possible entanglement and correlations in a tripartite system \cite{dur}. 
In quantum information, they have applications in e.g. quantum teleportation \cite{tele},
dense coding \cite{dense}, and
measurement-based quantum computing \cite{one-way}.
Previously, the GHZ-states have been demonstrated in for example superconducting qubits \cite{dicarlo,neeley}, and 
theoretically proposed in e.g. single-spin qubit systems \cite{ghzdot1,ghzdot2}. 

A promising realization for a quantum bit \cite{levy,Loss98} is the two-electron spin eigenstates in quantum dots (QD) \cite{Ashoori96,Reimann02,Saarikoski_RMP}. 
The universal set of quantum gates \cite{taylor2} for two spin singlet-triplet 
qubits has been demonstrated experimentally \cite{petta2, foletti, shulman}. In this architecture, the inter-qubit interactions can be implemented
using capacitative coupling that exploits the differences between the charge configurations of the singlet and triplet states
to generate entanglement between the qubits \cite{stepa,taylor}. The scheme, a two-qubit capacitatively coupling
CPHASE-gate, has been realized experimentally quite recently \cite{shulman,weperen}.

In this paper, we propose a generalization of the CPHASE-gate, a capacitative three-qubit gate that creates tripartite entanglement between singlet-triplet qubits. The qubits
are placed in the corners of a triangle. As they are evolved under exchange interaction, generated by electronically detuning the qubits,
they start to entangle and disentangle. We quantify the classes of entanglement generated in the gate and show that a GHZ-state can be obtained by the gate-operation.

Our method provides an efficient way for generating high tripartite entanglement, as the gate operation does not include multiple steps, but just detuning the three qubits
to the desired values of the exchange interaction. Our analysis can be used to determine the pulse sequences to be used in such gate for the creation of long-lasting GHZ-states, paving the way towards an experimental realization of these maximally entangled states and three-qubit interactions
in the singlet-triplet qubit architecture.

A system consisting of $N_q$ singlet-triplet qubits ($N=2N_q$ electrons and QDs) is modeled using an extended Hubbard model Hamiltonian,
\begin{eqnarray}\label{eq:ham}
H&=&\sum_{i=1}^{2N}E_ia^{\dagger}_ia_i\nonumber \\ &-&\sum_{i,j=1}^{2N}t_{ij}a_i^{\dagger}a_j 
+\sum_{i,j,k,l=1}^{2N}v_{ijkl}a_i^{\dagger}a_j^{\dagger}a_la_k,
\end{eqnarray}
where $E_i$ are the on site energies at each QD, $t_{ij}$ the tunneling matrix elements between dots, and $v_{ijkl}$ the Coulomb-interaction
matrix element. Here the indices $i,j,k,l$ refer to both the spin and spatial degrees of freedom.
$\{a^{\dagger}_i\}_{i \leq N}$ are the creation operators for the $\sigma_i=-1/2$ electrons at the sites from $1$ to $N$,
and $\{a^{\dagger}\}_{i>N}$ are those for the the $\sigma_i=+1/2$ electrons. Spin is conserved in the tunneling, i.e. $t_{ij}=\delta_{\sigma_i\sigma_j}t_{ij}$. 

In Eq. (\ref{eq:ham}), the Coulomb-interaction is long-range, $v_{ijkl}=\langle i|\langle j|\frac{C}{|\mathbf{r}_1-\mathbf{r}_2|}|l\rangle| k\rangle$.
With $\{|i\rangle\}$ being delta-functions in the Hubbard model, the $v_{ijkl}$-elements can be written as
\begin{equation}
v_{ijkl}=\delta_{ik}\delta_{jl}\left[(1-\delta_{ij})\frac{C}{|\mathbf{r}_i-\mathbf{r}_j|-d}+\delta_{ij}U\right].
\end{equation}
Here, $C=e^2/4\pi\epsilon_r\epsilon_0$ is the Coulomb-strength, $\mathbf{r}_i$ and $\mathbf{r}_j$ are the locations of the dots $i$ and $j$. $U$ is the on-site interaction between two electrons in the same QD and $d>0$ is an extra constant
conveying the fact that in truth the wave functions have finite widths. 

In the capacitative coupling, the tunneling between two $S-T_0$ qubits is usually negligible \cite{taylor,stepa,weperen,shulman}. Hence, in our model, the tunneling elements $t_{ij}$
are non-zero only between the two dots inside the qubits. The Hamiltonian of Eq. (\ref{eq:ham}) is diagonalized 
in the $S_z=0$ subspace (i.e. the number of both up and down electrons is $N_q$) to obtain the eigenstates of the system. 

The parameters $t_{ij}$, $U$ and $d$ can be fitted to exact diagonalization (ED) data \cite{leak} in order to produce realistic results.
We compare the Hubbard-results to a reference system of two capacitatively coupled $S-T_0$ qubits (four QDs)
modeled as parabolic potential wells.
The parabolic dot minima are located at the $x$-axis. The dot distances in the qubits are
$80$ nm, and the inter-qubit distance is $120$ nm. The parabolic confinement strength is $\hbar\omega_0=4$ meV. The GaAs-value of $\epsilon_r\approx12.7$ is used
for the permittivity. The many-body basis of Slater determinants
is created using the single-particle eigenstates of the system which are computed using the multi-center gaussian method \cite{requ}.
The many-body Hamiltonian is diagonalized using the Lanczos-method. See
\cite{leak} for more details on the ED-calculation. 

\begin{figure}[!ht]
\vspace{0.3cm}
\includegraphics[width=\columnwidth]{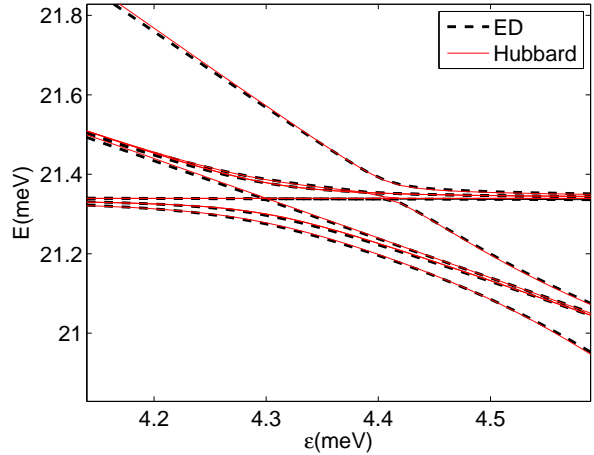}
\caption{(Color online) The lowest energies of a two-qubit system as function of
the detunings $\epsilon_1=\epsilon_2=\epsilon$. The thick black dashed line shows the ED-energies, and the
red line the Hubbard energies with the parameters $t_{ij}=27.8$ $\mu$eV, $U\approx3.472$ meV, and $d=0.43$ nm.}
\label{fig:ed_vs_hub}
\end{figure}

A good fit is obtained with the values $t_{ij}=27.8$ $\mu$eV, $U\approx3.472$ meV, and $d=0.43$ nm. The lowest energies (including the relevant two-qubit states
$|SS\rangle$, $|ST_0\rangle$, $|T_0S\rangle$, and $|T_0T_0\rangle$) can be seen in Fig. \ref{fig:ed_vs_hub}. as a function of the detunings $\epsilon_1$ and $\epsilon_2$.
($\epsilon_n$ is the difference of the on-site energies $E$ between the two dots of the qubit $n$). Here, $\epsilon_1=\epsilon_2=\epsilon$.
The energies computed with the two methods coincide almost exactly.
The obtained parameters were also tested in asymmetric detuning cases, $\epsilon_1\neq\epsilon_2$, and the fit was equally good there.
As the strength of the
capacitative interaction is determined by the energy differences of the two-qubit basis states, our Hubbard model should now describe the qubit-qubit interactions
realistically (at least with these used intra and inter-qubit dot distances).

We then move on to coupling three qubits to each other. In the three-qubit gate, the qubits are placed symmetrically at the corners
of an equilateral triangle (both the inner and outer dots of the qubits form an equilateral triangle). The distance of the qubits (i.e. the distance
of the inner dots) is $120$ nm, and the intra-qubit dot distance is $80$ nm, as in Fig. \ref{fig:ed_vs_hub}. The parameters of the Hubbard model correspond to $\hbar\omega_0=4$ meV confinement
in the dots, as in Fig. \ref{fig:ed_vs_hub}.

The three-tangle, $\tau_{123}$, measures the tripartite-entanglement in a three-qubit system \cite{dur,coff}. Writing an arbitrary state $|\psi\rangle$ of the three qubits as
$|\psi\rangle=\sum_{i,j,k=0}^1a_{ijk}|ijk\rangle$, with $|ijk\rangle=|i\rangle_1\otimes|j\rangle_2\otimes|k\rangle_3$, the three-tangle (for pure states) is given as
\begin{eqnarray}
\tau_{123}=&2&\left|\sum a_{ijk}a_{i'j'm}a_{npk'}a_{n'p'm'} \right. \nonumber \\
&\times&\left.\eta_{ii'}\eta_{jj'}\eta_{kk'}\eta_{mm'}\eta_{nn'}\eta_{pp'}\right|,\label{eq:tang}
\end{eqnarray}
where the sum goes from $0$ to $1$ for all indices, and $\eta_{01}=1$, $\eta_{10}=-1$, and $\eta_{ii}=0$.
The value of $\tau_{123}$ is between $0$ and $1$, and it is maximal for the GHZ-states. Conversely, the W-class states have zero tangle.
The entanglement in a subsystem of two qubits can be measured by pairwise concurrences.
The pairwise concurrence $C_{12}$ of qubits $1$ and $2$ can be computed from the reduced density matrix of the pair (see e.g. \cite{coff} for the details).
The W-states maximize all three pairwise concurrences, so that $C_{12}^2+C_{13}^2+C_{23}^2=4/3$. In the GHZ-states, the pairwise concurrences are zero \cite{coff}. 

We evolve
the qubits under exchange (the detunings $\epsilon_i$ are held constant), so that the $S$ and $T_0$-states have differing charge configurations, causing the qubits to entangle with each other.
The time evolution of the full six-electron wave function $|\Psi\rangle$ is computed as $|\Psi(t+\Delta t)\rangle=\exp\left(-i\Delta tH(t)/\hbar\right)|\Psi(t)\rangle$, where
$H(t)$ is the Hamiltonian (\ref{eq:ham}) at time $t$.
We project the wave function $|\Psi\rangle$ onto the three-qubit basis $\{|ijk\rangle\}_{i,j,k=0,1}$,
and compute the 3-tangle $\tau_{123}$ and the pairwise concurrences at each time step to study the entanglement properties of the system. Here, we write
the singlet state $S$ as $0$ and the triplet $T_0$ as $1$. 

\begin{figure}[!ht]
\vspace{0.3cm}
\includegraphics[width=\columnwidth]{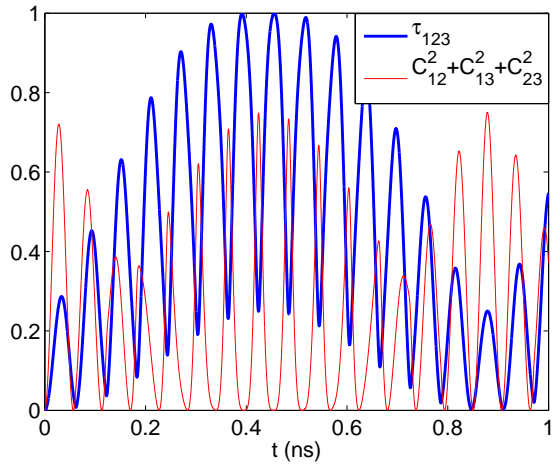}
\caption{(Color online) The evolution of the concurrences and the three-tangle with the detunings $\epsilon_1=\epsilon_2=\epsilon_3=\epsilon=3.9$ meV. The thick
blue line shows the three-tangle and the red line the squared sum of the pairwise concurrences. At $t=0$, all qubits are initiated
in the $xy$-plane of the Bloch sphere. The qubits are then let to evolve, and the concurrences and the three-tangle are computed at each time step.}
\label{fig:det_40}
\end{figure}

Fig. \ref{fig:det_40}. shows the evolution
of the concurrences in the case with the detunings held at $\epsilon_1=\epsilon_2=\epsilon_3=\epsilon=3.9$ meV (the further away dots are detuned to low potential). The system is initiated in a product state
$|\Psi\rangle=|\psi\rangle_1\otimes|\psi\rangle_2\otimes|\psi\rangle_3$, with $|\psi\rangle_i=\frac{1}{\sqrt{2}}(|0\rangle+e^{i\phi_i}|1\rangle)$, with random phases $\phi_i$. 
Both the pairwise concurrences and the three-tangle start to oscillate
as the qubits are evolved. Both oscillations follow a similar modulated form, with fast carrier wave-like oscillations.

At around $t=0.46$ ns, $\tau_{123}$
assumes the value of $1$ (the exact numerical maximum being $\tau_{123}=0.9999987$ in this simulation). At the same time, the pairwise concurrences are zero.
These are both characteristics of GHZ-states. Indeed, the state is found to be a GHZ-state $|\overline{GHZ}\rangle$ that
can be obtained from the more 'typical' GHZ-states $|GHZ^{\alpha}\rangle=\frac{1}{\sqrt{2}}(|000\rangle+e^{i\alpha}|111\rangle)$ by single-qubit rotations.
Generally, these one-qubit operations are given by rotating all three qubits the angle of $\pi/4$ around an axis $\hat{\mathbf{n}}(\varphi)=\cos(\varphi)\hat{\mathbf{x}}+\sin(\varphi)\hat{\mathbf{y}}$ that lies in the $xy$-plane of the Bloch sphere.
For example, the maximal tangle in Fig. \ref{fig:det_40}. corresponds to rotating the qubits $\pi/4$ around the $x$-axis ($\varphi=0$) so that
$|\overline{GHZ}\rangle=e^{i\pi/4\cdotp(\mathbf{X}\otimes\mathbf{I}\otimes\mathbf{I}+\mathbf{I}\otimes\mathbf{X}\otimes\mathbf{I}+\mathbf{I}\otimes\mathbf{I}\otimes\mathbf{X})}|GHZ^{\pi/2}\rangle$, where
$\mathbf{I}$ is the indentity, and $\mathbf{X}$ is the $x$-Pauli matrix.

\begin{figure}[!ht]
\vspace{0.3cm}
\includegraphics[width=\columnwidth]{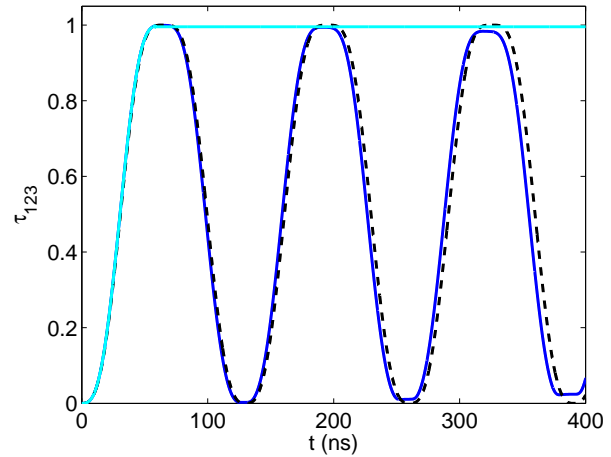}
\caption{(Color online) The evolution of the three-tangle with the detunings $\epsilon=3.5$ meV. At $t=0$, the qubits are initiated
in the $xy$-plane of the Bloch sphere. The qubits are then let to evolve, and the three-tangle is computed at each time step. The
blue line (dark solid line) shows the three-tangle computed by projecting the wave function onto the qubit-basis.  
The dashed black line shows the approximated tangle according to Eq. (\ref{eq:tau2}). The cyan line (light solid line) shows the long-lasting $|\overline{GHZ}\rangle$-
state obtained by switching off the detunings when $\tau_{123}$ reaches its maximum. Here, the detunings are decreased to zero in a time of $0.5$ ns, and the
system is then let to evolve again.}
\label{fig:det_355}
\end{figure}

The value of the detuning affects the operation of the three-qubit entangling procedure. Small values of $\epsilon$ lead to small differences in the charge distributions of the qubit states and hence to slower operation.
However, there are also qualitative differences between the concurrence oscillations with different detunings.
The frequency of the carrier oscillations
of $\tau_{123}$ varies in a quite complex manner compared to the modulating envelope's frequency as a function of the detuning $\epsilon$. With small or very high values of the detunings ($\epsilon<3.6$ meV or $\epsilon>4.4$ meV),
the modulation is very slow compared to the carrier oscillations. This is demonstrated in Fig. \ref{fig:det_355}. as the blue line (dark solid line). Here, $\epsilon=3.5$ meV. It almost seems that there is
no modulation at all, but just $\tau_{123}$ oscillating between $0$ and $1$. In contrast, with the intermediate detuning values ($3.6$ meV $\leq\epsilon\leq4.4$ meV),
the oscillations can become quite complicated, with two or three modulating envelopes on top each other. 

The behavior of the oscillations can be explained by studying the energy differences of the qubit basis states. As our system is perfectly symmetrical
with respect to the three qubits, the states $|100\rangle$, $|010\rangle$, and $|001\rangle$ are degenerate. The same applies to $|110\rangle$, $|101\rangle$, and $|011\rangle$.
The energies of the three-qubit basis states can thus be divided into four categories, $E_1$ corresponding to $|000\rangle$, $E_2$ corresponding to $|100\rangle$ etc, 
$E_3$ ($|110\rangle$ etc), and $E_4$ corresponding to $|111\rangle$.

The qubits are initiated in the $xy$-plane (for simplicity, now in the states $|\psi\rangle_i=\frac{1}{\sqrt{2}}(|0\rangle+|1\rangle),i=1,2,3$. A similar
result applies for the general case). The full six-body wave-function is thus $|\Psi\rangle=\frac{1}{\sqrt{8}}\sum_{i,j,k=0}^1|ijk\rangle$. As the qubits
are now evolved with constant detunings, $\epsilon_1=\epsilon_2=\epsilon_3=\epsilon$, the gate functions as a three-body version of the CPHASE-gate, i.e. each term $|ijk\rangle$ obtains a phase-factor corresponding to its energy. Inserting the phases into Eq. (\ref{eq:tang}), yields
an analytic formula for the evolution of the three-tangle,
\begin{eqnarray}
\tau_{123}(t)&=&\frac{1}{16}\left|e^{-2\delta_2it/\hbar}-6e^{-\delta_2it/\hbar}+4e^{-\delta_1it/\hbar}\right. \nonumber\\
&+&\left.4e^{-(\delta_2-\delta_1)it/\hbar}-3\right|. \label{eq:tau1}
\end{eqnarray}
Here, $t$ is the time, and we have written: $E_2-E_1=\Delta$, $E_3-E_2=\Delta+\delta_1$, and $E_4-E_3=\Delta+\delta_2$.
Eq. (\ref{eq:tau1}) is indeed found to produce the same exact tangle-oscillations observed in the simulations. The frequencies of the modulation and the carrier oscillations
(e.g in Fig. \ref{fig:det_40}) are given by $\delta_1$, and $\delta_2-\delta_1$, respectively. However, the phase of the modulation varies, and in some cases,  
for example, with $\epsilon=3.8$ meV, there are more than one modulation envelopes on top of each other. 

In the cases of low and very high detunings, Eq. (\ref{eq:tau1})
can be simplified. The anti-crossing region of the singlet charge-states is located between $\epsilon=3.6$ meV
and $\epsilon=4.4$ meV in this three-qubit system. Before the anti-crossing ($\epsilon<3.6$ meV), the singlets are in the $(1,1)$-configuration in all of the
qubit basis states. After it ($\epsilon>4.4$ meV), all the singlets have undergone the transition to the doubly-occupied $(0,2)$-states.
Outside of the anti-crossing region, it turns out that $(E_4-E_3)-(E_2-E_1)\approx2[(E_3-E_2)-(E_2-E_1)]$, i.e. $\delta_2\approx2\delta_1$.

With $\delta_2=2\delta_1$, Eq. (\ref{eq:tau1}) simplifies to
\begin{equation} \label{eq:tau2}
\tau_{123}(t)=\frac{1}{4}\sqrt{\left[1-\cos(t\delta_1 /\hbar)\right]^3\left[5+3\cos(t\delta_1 /\hbar)\right]}.
\end{equation}
Eq. (\ref{eq:tau2}) oscillates between $0$ and $1$ with the frequency of $f=\delta_1/2\hbar\pi$.
We find that the corresponding oscillation frequency can also be determined by studying the pairwise concurrences, e.g. $C_{12}$ in the case when
only the qubits $1$ and $2$ are detuned, so that $\epsilon_1=\epsilon_2=\epsilon$, $\epsilon_3=0$. In this case, the pairwise concurrence
is found to oscillate between $0$ and $1$ as $C_{12}(t)=\frac{1}{2}\sqrt{2-2\cos(2\pi ft)}$, with the same frequency $f$ as in Eq. (\ref{eq:tau2}).

Eq. (\ref{eq:tau2}) thus allows one to predict the moment when the $|\overline{GHZ}\rangle$-state is obtained with the gate operation.
It fits quite well with the slowly modulating tangle shown in Fig \ref{fig:det_355}, especially in the beginning of the oscillation. With longer gate-operation,
the tangle given by Eq. (\ref{eq:tau2}) starts to deviate more from the simulated one, as the modulation still exists
(the further away the detunings are from the anti-crossing area, the better $\delta_2\approx2\delta_1$ holds).

In the anti-crossing area,
the singlet in e.g. $|110\rangle$ may be fully in the $(0,2)$-state, while those of $|100\rangle$ are still in a superposition of $S(1,1)$ and $S(0,2)$  due to higher Coulomb-repulsion between the inner dots
in $|110\rangle$. In this case, $\delta_2>2\delta_1$, and Eq. (\ref{eq:tau2}) no longer holds. Instead, $\tau_{123}$ evolves in a complex manner with
carrier and envelope oscillations, as seen in Fig. \ref{fig:det_40}. 

Next, we discuss using the three-qubit gate to create a 'long-lasting' GHZ-state. The three-tangle
keeps oscillating as long as the exchange interaction in the qubits is non-zero. In principle,
turning the detunings off the exact moment when $\tau_{123}$ has reached its maximum, yields a stationary $|\overline{GHZ}\rangle$-state (apart from decoherence effects \cite{khaet,folk,cnoise} that
are not currently included in our model). However, due to the effect of the charge state leakage \cite{leak}, the detuning must be turned off gradually. However, if the detuning is decreased slow enough to ensure an adiabatic passage
from $S(1,1)$ to $S(0,2)$, the qubits keep interacting during the decrease, and the state is no longer the maximally entangled GHZ-state. 

With small detunings and slower gate operation (like in Fig. \ref{fig:det_355}),
the decrease times can be longer. In addition, smaller detunings can be turned off faster while still retaining adiabaticity. Also,
the less chaotic oscillations with small detuning allow better predictability of the location of the $\tau_{123}$ maxima, via Eq. (\ref{eq:tau2}). In experiments, a longer gate-operation time means more decoherence.
Hence, a compromise between fast operation and easy state preparation should be done in possible experimental realizations of this entangling scheme.

The procedure is demonstrated in Fig. \ref{fig:det_355}. as the cyan (solid light) line. The qubits are again 
initiated in the $xy$-plane, and let to evolve under exchange. The detunings are $\epsilon=3.5$ meV. When $\tau_{123}$ reaches its maximum, the detunings are
decreased to zero linearly during a time of $0.5$ ns. When the detunings have reached zero, the system is let to evolve again. As seen in the figure,
the three-tangle stays at its maximum after the detuning sweep.

In our discussion, the three-qubit system has thus far been perfectly symmetrical. Lastly, we briefly discuss the effect of asymmetry on the entanglement properties of the system.
The symmetry was broken by by adding a small random vector $\mathbf{d}_i$ to the location
of each dot, $\tilde{\mathbf{r}}_i=\mathbf{r}_i+\mathbf{d}_i$ or by using asymmetric detunings, $\epsilon_1=\epsilon$, $\epsilon_2=\epsilon+\delta\epsilon$, $\epsilon_3=\epsilon-\delta\epsilon$.
The larger the asymmetry, the more chaotic the concurrence oscillations become, as the energies of the qubit states contain less symmetry ($|001\rangle$ and $|010\rangle$ are no longer degenerate, and so on).
A maximal $\tau_{123}=1$ GHZ-state is still obtained even with very large dislocations or detuning differences (e. g. $|\mathbf{d}_i|=10$ nm, $\delta\epsilon=0.5$ meV).
It should be noted that the pairwise concurrences seem to obtain larger values (up to $C_{12}^2+C_{13}^2+C_{23}^2\approx 1.25$) in the asymmetrical cases.

In conclusion, we have proposed an entangling three-qubit gate based on the capacitative coupling of singlet-triplet qubits. It provides a simple and efficient method for generating maximally
entangled tripartite states in the singlet-triplet qubit architecture.
We analyze the gate operation using an accurate microscopic model, and find that a GHZ-state can be generated. We derive an analytical
formula for the evolution of three-body entanglement valid for small and very high detuning values. Using the formula and our analysis, one can determine
the detuning pulse sequences to be used for generating the GHZ-states. We also show that by turning off the detunings at the right phase of the oscillation,
one can create a long-lasting GHZ-state of maximal three-qubit entanglement.

\begin{acknowledgments}
We thank Mikko Ervasti and Zheyong Fan for comments and suggestions.
This research has been supported by the Academy of Finland through its Centres of Excellence Program (project no. 251748).
\end{acknowledgments}
\bibliography{lagrange.bib}

\end{document}